\documentclass[twocolumn,showpacs,preprintnumbers,amsmath,amssymb,prl]{revtex4-1}
\usepackage{tikz}
\usetikzlibrary{calc,decorations.markings}
\usepackage{epsfig}
\usepackage{graphicx}
\usepackage{rotating}
\usepackage{amssymb}
\usepackage{amsmath}
\usepackage{amsthm}
\usepackage{float}   
\usepackage[active]{srcltx}
\usepackage{url}
\usepackage{hyperref}
\usepackage{centernot}
\usepackage{dsfont,bm}
\usepackage{xcolor}
\usepackage{pgfplots}

\newtheorem{prop}{Proposition}
\newtheorem{lem}{Lemma}
\hypersetup{
	colorlinks   = true, 
	urlcolor     = blue, 
	linkcolor    = blue, 
	citecolor   = blue 
}

\graphicspath{{figs/}}

\begin{document}

\title{Single Qubit Error Mitigation by Simulating Non-Markovian Dynamics}

\author{Mirko Rossini, Dominik Maile, Joachim Ankerhold}
\affiliation{Institute for Complex Quantum Systems and IQST, Ulm University - Albert-Einstein-Allee 11, D-89069 Ulm, Germany}
\author{Brecht I. C Donvil}
\email{brecht.donvil@uni-ulm.de}
\affiliation{Institute for Complex Quantum Systems and IQST, Ulm University - Albert-Einstein-Allee 11, D-89069 Ulm, Germany}

\begin{abstract}
Quantum simulation is a powerful tool to study the properties of quantum systems. The dynamics of open quantum systems are often described by Completely Positive (CP) maps, for which several quantum simulation schemes exist. We present a simulation scheme for open qubit dynamics described by a larger class of maps: the general dynamical maps which are linear, hermitian preserving and trace preserving but not necessarily positivity preserving. The latter suggests an underlying system-reservoir model where both are entangled and thus non-Markovian qubit dynamics. Such maps also come about as the inverse of CP maps.  We illustrate our simulation scheme on an IBM quantum processor by showing that we can recover the initial state of a Lindblad evolution. This paves the way for a novel form of quantum error mitigation.  Our scheme only requires one ancilla qubit as an overhead and a small number of one and two qubit gates.
\end{abstract}
\maketitle


{\em Introduction.-} Quantum computing has created a computational paradigm that may lead to the development of new and powerful solutions to computational tasks. 
A prominent application of digital quantum computers is their ability to simulate other quantum systems, already on the level of noisy intermediate scale (NISQ) quantum platforms \cite{GeAs2014}.

Although there already exists a wide range of quantum simulation methods for closed quantum systems, see e.g. \cite{Llo1996,BeGr2006,Chi2009,WiDo2011},
the simulation of an open quantum system is a more arduous task. Since it is often not possible to simulate the complete system plus environment dynamics, simulation methods mainly focus on realizing the reduced effective dynamics of the open quantum system.

\begin{figure}[tt]
\centering
\includegraphics[scale=1.1]{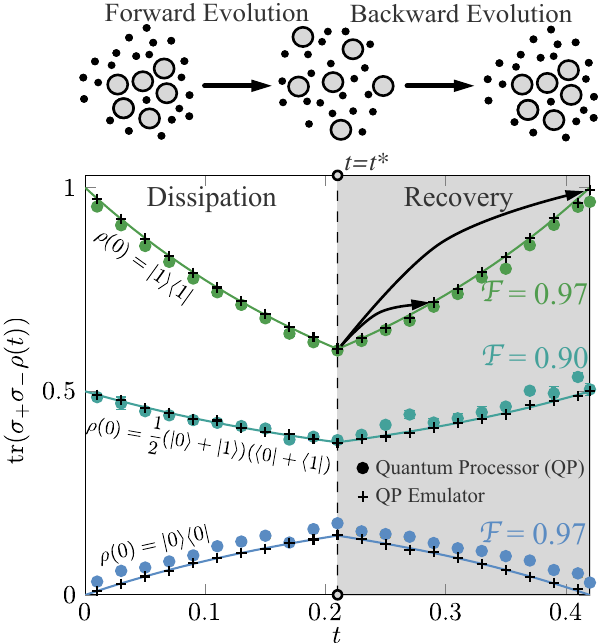}
\caption{(Top) Reversing the direction of time of a dissipative system brings it back to its initial state. (Bottom) We recover the initial state of a Lindblad evolution \eqref{eq:therm_for} by simulating its time-reversed evolution $\frac{d}{dt}\rho_t=-\mathcal{L}_t(\rho_t)$, for different initial states. For $t\leq t^*$ the points show the forward evolution simulated on an IBM quantum computer, for $t>t^*$ the backward evolution. The lines are the numerical integration of the master equations \eqref{eq:therm_for} and their time-reversed versions. $\mathcal{F}$ gives the fidelity between the final recovered state and the initial state. The parameters of the master equations are $\beta=\omega=\gamma=1$. }
\label{fig:recovery}
\end{figure}

To find reduced descriptions for the evolution of open quantum systems, one typically assumes that system and environment are initially in a product state. In this case, the evolution is guaranteed to be described by a Completely Positive (CP) map \cite{Rivas} acting on the initial system state, which can be obtained by resorting to numerical methods (such as \cite{StrKi2018,SuEi014,PrCh2010,TaKu1989,XuYa2022}) or perturbative schemes \cite{BrPe2020}. 
A CP map is said to be CP-divisible if it can be divided into small arbitrary parts that are themselves CP. In this case, the evolution of the system can be described by the Lindblad-Gorini-Kossakowski-Sudarshan equation \cite{Li1976,Gorini1976}. When a significant amount of entanglement accumulates between the system and the environment, the use of CP maps no longer makes sense \cite{Rivas2020}. 

Quantum simulation methods for Lindblad-like dynamics have been extensively studied \cite{BaCh2001,LloVio2001,WeMu2010,WaAs2011,KliBa2011,BaKl2012} and experimentally implemented \cite{Schindler2013}. Recently the authors of \cite{GuLi2023} simulated Lindbladian dynamics on an IBM quantum processor by exploiting error mitigation. An efficient simulation scheme for any CP qubit map was developed by \cite{WaBe2013}, based on the results of \cite{KingRus2001,RusSza2002}. The scheme by \cite{WaBe2013} uses a single ancillary qubit, single qubit rotations and CNOT gates and was experimentally realised on a superconducting circuit \cite{HaCa2021}. 

In this Letter we propose a novel simulation scheme for general qubit dynamical maps going beyond CP divisible maps \cite{KingRus2001,RusSza2002,WaBe2013}. General dynamical maps are trace-preserving and hermicity-preserving, but not necessarily positivity-preserving. They naturally arise as the inverse of CP maps, and thus the ability to simulate them allows to revert noise effects on the system. In Fig. \ref{fig:recovery} we recover several initial states of a qubit disturbed by a thermal Lindblad equation on an IBM quantum processor. General dynamical maps also arise in the context of general time-local master equations in which the initial state is {\em not} assumed to be in a product state, but has already created some system-bath entanglement. Their derivation is based on tracing out environmental degrees of freedom in Gaussian boson \cite{Feynman1963,KaGr1997}, fermion \cite{Tu2008,DoGo2020} or other exactly solvable models \cite{JoQu1994}, or via time convolutionless perturbation theory methods \cite{HaFu1977,BrPe2020}.

For our proposed scheme, we exploit the fact that general dynamical maps acting on finite dimensional systems can always be decomposed as the difference of two CP maps \cite{SweSa2016,SuppQubitMaps}. We demonstrate that the decomposition can be brought into a suitable form for quantum simulations, namely, as a weighted difference of two completely positive trace-preserving (CPTP) maps.

Another, more general avenue for the simulation of open quantum systems on a quantum computer are collision methods \cite{CiSa2022,CaDe2021}. Here the open quantum system repeatedly interacts or "collides" with ancillary systems. Non-Markovian dynamics can be implemented by allowing the ancillas to interact amongst themselves in between system-ancilla collisions \cite{McPa2014,LoCi2016,KreKi2016}. Such models were successfully implemented on IBM quantum processors \cite{GaRoMa2020}. Other methods for quantum simulation of open quantum systems have been proposed in the context of non-equilibrium systems \cite{LamLaw2018} and quantum thermodynamics \cite{ChenLiDong2021}. 

In contrast to collision-based methods such as \cite{McPa2014,LoCi2016,KreKi2016}, the main advantage of the algorithm we present here is its problem-agnostic applicability. No specific design is required for each dynamics one aims to simulate.
In fact, we are able to simulate the dynamics of a given system from any point in time to any later time, whereas collision-based methods require the protocol to always start from the initial time of the evolution and rely on the Trotterization of the dynamics. As an additional benefit, our proposed scheme is resource efficient as the computational overhead does not grow with the simulated time. 

We illustrate these features by implementing two paradigmatic examples on IBM quantum processors which demonstrate the ability to simulate the time evolution from any intermediate point in time, even when the evolution map is not CP and the ability to recover the initial state of a Lindbladian evolution \cite{DoMu2022arx}, see Fig. \ref{fig:recovery}.

This last example shows that one of the promising applications of the new scheme is to perform error mitigation on a NISQ platform by recovering the typically unknown undisturbed initial state.

\begin{figure}
\centering
\includegraphics[scale=1]{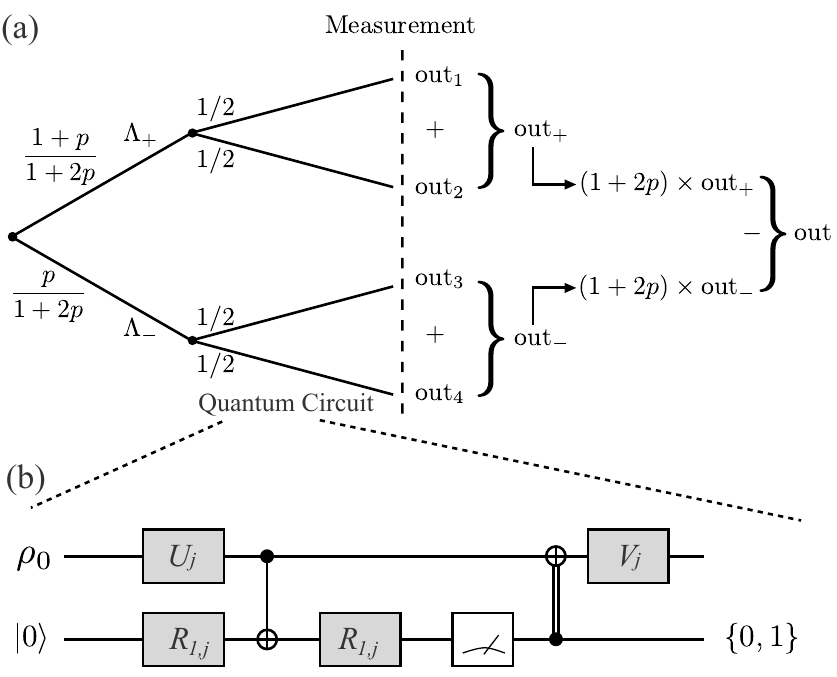}
\caption{(a) Representation of the algorithm simulating general dynamical maps decomposed as the weighted difference of two CPTP maps \ref{eq:wittstock-paulsen2}. At each branching point a choice is made with a classical random number generator with the indicated probability. At the dashed line, the measurement of some observable is performed. Finally, the outcomes are rescaled and subtracted from one another. (b) The circuit by \cite{WaBe2013} to realise the extremal maps \eqref{eq:explicitt-kraus}.}.
\label{fig:scheme}
\end{figure}

{\em Theoretical framework.-} A finite dimensional linear map $\Lambda$ is CP if and only if its action on a state $\rho$ can be written in terms of a set of matrices $\{K_j\}_j$, often referred to as Kraus operators: $
\Lambda(\rho) = \sum_j K_j \rho K_j^\dagger$.
The map is trace preserving iff. $\sum_j K_j^\dagger K_j = \mathbb{I}$, where $\mathbb{I}$ is the identity on the appropriate Hilbert space, see e.g. \cite{BrPe2020,Rivas}. The results of \cite{KingRus2001,RusSza2002} prove that any CPTP qubit map $\Lambda$ is equal to the convex sum of two extremal CP maps $\Lambda_1$ and $\Lambda_2$ that are both realised by a pair of Kraus operators. Concretely, they showed that for every CPTP qubit map  $\Lambda$ there exist two pairs of unitaries $U_{j}$, $V_{j}$ and two pairs of Kraus operators $F_{i,j}$ (with $i,j \in \{1,2\}$) defining the extremal maps
\begin{equation}
\Lambda_j(\rho) = U_j \left(\sum_{i=1}^2 F_{i,j} (V_j\rho V_j^\dagger) F_{i,j}^\dagger\right) U_j^\dagger
\end{equation}
such that 
\begin{align}\label{eq:explicitt-kraus}
\Lambda(\rho) = \frac{1}{2}\Lambda_1(\rho)+\frac{1}{2}\Lambda_2(\rho).
\end{align}  
The authors of \cite{WaBe2013} devised a simple circuit shown in Fig. \ref{fig:scheme} (b) to realise the action of the $\Lambda_j$ using just one ancillary qubit and CNOT gates.

General dynamical maps are linear, trace-preserving and self-adjoint but not necessarily positivity preserving. For finite dimensional systems such maps can always be written as the difference of two CP maps \cite{SuppQubitMaps}
\begin{align}\label{eq:wittstock-paulsen}
\Sigma(\rho) = \Lambda_+(\rho)-\Lambda_-(\rho) =  \sum_j K_{j}\rho K_{j}^\dagger -\sum_j M_{j}\rho M_{j}^\dagger.
\end{align}
The map $\Sigma$ is trace preserving under the condition $\sum_j K_j^\dagger K_j -\sum_j M_j^\dagger M_j  = \mathbb{I}$. Since $\Lambda_\pm$ are bounded, there exists a positive number $p$ such that $\sum_j M_j^\dagger M_j  \leq p\, \mathbb{I}$. We define the semi-positive definite operator $D =\sqrt{p \,\mathbb{I}-\sum_j M_j^\dagger M_j}$ and write
\begin{align}\label{eq:wittstock-paulsen2}
\Sigma(\rho)= (1+p) \Lambda^*_+ - p\Lambda^*_-,
\end{align}
where $\Lambda^*_+=\frac{\sum_j K_{j}\rho K_{j}^\dagger +D\rho D^\dagger}{1+p}$ and $\Lambda^*_-= \frac{\sum_j M_{j}\rho M_{j}^\dagger+ D\rho D^\dagger}{p}$. It is straightforward to check that both maps $\Lambda^*_\pm$ are trace preserving and CP. 
The above equation is our first main result. It shows that any general dynamical map $\Sigma$ can be decomposed as the weighted difference of two CPTP maps. The above decomposition is an alternative to the decomposition in terms of CPTP maps by \cite{SuSha2003} which involves applying positive, non-unitary transformations to the quantum state. The latter makes our decomposition more attuned for simulating the action of $\Sigma$ on an experimental platform.

{\em Algorithm and circuit schemes.-} The simulation scheme we propose for general dynamical maps is illustrated in Fig. \ref{fig:scheme}(a). First, a classical random number generator is used to choose the branch $\Lambda_+^*$ or $\Lambda_-^*$ corresponding to equation \eqref{eq:wittstock-paulsen2}, with probabilities $\frac{1+p}{1+2p}$ and $\frac{p}{1+2p}$, respectively. Then one of the two extremal maps \eqref{eq:explicitt-kraus} is selected with probability 1/2 and realized by the circuit representation for the extremal maps of \cite{WaBe2013} shown in Fig. \ref{fig:scheme}(b).  Next, a measurement of an observable is performed and the outcomes within the plus and minus branch are summed. Finally, the measurement result is rescaled by $1+2p$ to restore normalization, and the results of both branches are subtracted. 

The scheme depicted in Fig. \ref{fig:scheme} can be straightforwardly be implemented on a quantum computational platform. Particularly, we use the Ehningen IBM quantum device. Generating the dynamics of a system using the algorithm described above requires eight different quantum circuits as shown in Fig.\ref{fig:scheme}(b). Every circuit requires single-qubit unitary gates $U_j$, $V_j$, $R_{1,j}$ and $R_{2,j}$, which we construct explicitly in \cite{SuppQubitMaps}, are realized via a universal set of single-qubit gates. Beyond single-qubit unitary gates, CNOT gates and a measurement operation on the ancilla are performed. With this circuit representation any single qubit map $\Lambda$ can be simulated with an error $\leq \varepsilon$ using a computer time of $O(\textrm{polylog}(1/\varepsilon))$ \cite{WaBe2013}.

In order to minimize the noise effects of the quantum device, we implemented standard methods of quantum error mitigation \cite{SuppQubitMaps}. We select the qubits and connections on the platform showing the least error rate for each circuit implementation and optimise the specific gate protocol to minimise the number CX gates, being most prone to generate errors. We make use of readout error mitigation with an exploratory run on the device to uncover its systematic readout error and apply this to correct the measurement results.

\begin{figure}[b!]
\centering
\includegraphics[scale=1.3]{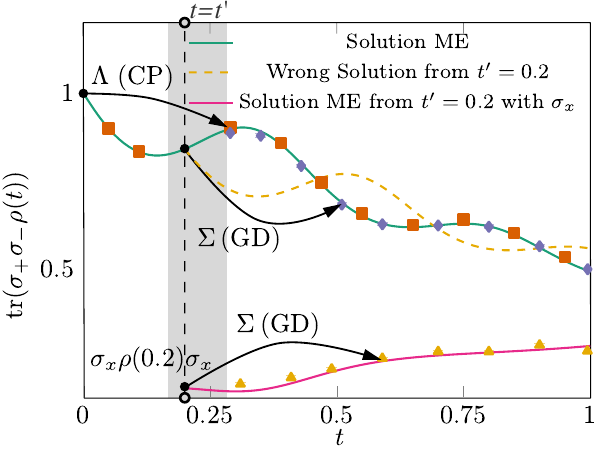}
\caption{Excited state population according to solutions of \eqref{eq:gen_time_loc} (lines) and IBM simulations (symbols), for an initial state $\rho_0= (1+\sigma_z)/2$. In the shaded region all weights are negative, so the evolution starting from there is ensured to be non-positivity preserving. The dynamical map $\Lambda_{t, 0}$ is CP, while starting at a later time $t'=0.2$, the map $\Sigma\equiv \Lambda_{t, t'}$ is only ensured to be a general dynamical map. The (yellow) line shows the evolution of the state from time $t'=0.2$ when forgetting about the past interaction with the environment.
 The purple line (green triangles) show the evolution from $t'=0.20$ after the unitary transformation $\sigma_x$ was applied to the state.}
\label{fig:sol}
\end{figure}

The data points in Figs. \ref{fig:recovery} and \ref{fig:sol} are each averaged over ten runs of each 10000 shots, i.e. 10000 circuits are implemented according to the probability distribution in Fig. \ref{fig:scheme}(a). Errors bars are within the size of the data points. Therefore, the final infidelity is mostly due to systematic errors in the quantum gates and the measurement scheme within a specific circuit calibration \footnote{These recalibrations are done on a daily basis by the IBM staff, see \href{https://quantum-computing.ibm.com/admin/docs/admin/calibration-jobs}{https://quantum-computing.ibm.com/admin/docs/admin/calibration-jobs}, accessed on the 6th of March 2023.}.  

{\em Simulating General Time Local Master Equations.-}
General trace-preserving time-local master equations are of the form 
\begin{align}\label{eq:gen_time_loc}
&\frac{d}{dt}\rho_t  =\mathcal{L}_t(\rho_t)\nonumber\\&
=-i [H_t,\rho_t]+\sum_k \Gamma_{k,t} (L_k\rho_tL_k^\dagger-\frac{1}{2}\{L_k^\dagger L_k,\rho_t\}),
\end{align}
where the $L_k$ are operators and the $\Gamma_k(t)$ are scalar weight functions.
The above equation has the appearance of a Lindblad equation except for the fact that the weight functions $\Gamma_k(t)$ are not assumed to be positive definite. General time-local master equations describe the evolution of a wide class of open quantum systems, as they can be derived from the Nakajima-Zwanzig equation \cite{ZwaR2001} when its solution has an inverse that exists during a finite time interval \cite{VstV1973,GrTaHa1977,vaWoLe1995,AnCrHa2007,ChKo2010}.
For an initial condition $\rho_0$ the formal solution of \eqref{eq:gen_time_loc} is $\rho_t=\Lambda_{t,0}(\rho_0)= T\exp\left(\int_0^tds\, \mathcal{L}_s\right)\rho_0$, where the map $\Lambda_{t,0}$ is guaranteed to be CP if the underlying system-environment model is in a product state.

The master equation \eqref{eq:gen_time_loc} generates maps that satisfy the semi-group property: For $\Lambda_{t,s}$ being the map that evolves a state from time $s$ to time $t$, then $\Lambda_{t,s}=\Lambda_{t,u}\Lambda_{u,s}$ for $t\geq u\geq s$. This property is very convenient
since we can split up the evolution into smaller segments that evolve the density matrix from one time to the next.  
However, complete positivity of $\Lambda_{t,0}$ does {\em not} necessarily guarantee that all intermediate maps $\Lambda_{s,u}$ are CP. In fact, if this is the case,  all weights $\Gamma_{k,t}$ are positive definite and \eqref{eq:gen_time_loc} reduces to the conventional Lindblad form.

If intermediate  $\Lambda_{u,s}$ are not positivity preserving, this implies that not all quantum states are mapped into quantum states, i.e. some quantum states are mapped into operators with negative eigenvalues. Indeed, weight factors $\Gamma_{k,t}$ taking negative values capture an underlying system-environment model with meaningful entanglement built up between them. In this case, the reduced system state operator at an instant of time is no longer sufficient to describe the subsequent time evolution, as one requires knowledge of the history of the system-environment interaction.
 
\begin{figure}
\centering
\includegraphics[scale=1.2]{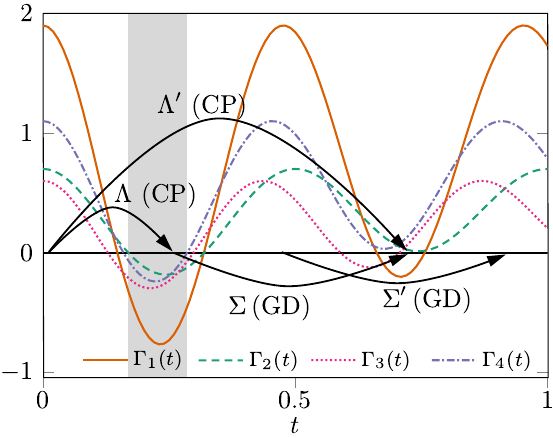}
\caption{Weight functions of the time local master equation \eqref{eq:gen_time_loc}. 
The rates are of the form $\Gamma_j(t)=a_j \exp(-t)(b_j-\sin^2(c_j\pi t))-d_j$, with $a_2=a_3=1$, $a_1= 3$ and $a_4 =1.5$; $b_1=4.5$, $b_2=3.5$, $b_3=1$ and $b_4=1.5$; $c_1=2$, $c_2=2$, $c_3=2.3$ and $c_4=2.2$, $d_1=d_2= 2.6$ and $d_3=d_4 = 0.4$}
\label{fig:rates}
\end{figure}

To illustrate this, we consider a qubit master equation with four operators and their respective weight functions, i.e.\
$L_1=\sigma_-$, $L_2=\sigma_+$, $L_3=\tau_-$ and $L_4=\tau_+$ with $\sigma_\pm$ being the raising and lowering operators of $\sigma_z$ and $\tau_\pm$ of $\sigma_x$, respectively.
As weight factors we choose a typical non-Markovian model with oscillations to negative values that exponentially decay, which mimics resonance with an environmental mode \cite{BrLa2016}. Figure \ref{fig:rates} displays these weight functions, where the grey zone indicates the times at which they are all negative. If the evolution of the density according to (\ref{eq:gen_time_loc}) starts in this time interval, for short times it will {\em not} be positivity preserving. Therefore, the solution of the master equation from these times has to be described within the above framework of general dynamical map.

The excited state population according to \eqref{eq:gen_time_loc} is shown in Fig. \ref{fig:sol}. Results obtained with the CP map $\Lambda_{t, 0}[\rho_0]$ with $\rho_0= (1+\sigma_z)/2$ are shown as (green) full line for a direct integration of (\ref{eq:gen_time_loc}) together with those from simulations on the IBM device (squares) following the recipe outlined above. Starting  at the intermediate time $t'=0.2$ in the grey area of Fig.~\ref{fig:rates}, diamonds display IBM simulations with the general dynamical map $\Sigma\equiv \Lambda_{t, t'}[\rho_{t'}]$. Since at $t'$ all weight functions are negative definite, the solution (for short times, at least) is not CP. In contrast, when forgetting about the past interaction with the environment according to an evolution with $\Lambda_{t-t', 0}[\rho_{t'}]$
(dashed yellow), the correct dynamics of (\ref{eq:gen_time_loc}) is not recovered.

Having access to the intermediate evolution maps starting from $t'>0$ has great advantages. For example, we are able to evolve a state from $0$ to $t'$, perform a quantum operation on it and then evolve it further with a completely bounded evolution. 
The (pink) full line in Fig. \ref{fig:sol} displays this situation after applying the unitary transformation $\sigma_x$ to the state at $t'=0.2$, while the (green) triangles show the corresponding IBM simulations of the general dynamical map according to the new scheme.

{\em Quantum State Recovery.-} Another intriguing consequence of the new simulation method is that one can recover the initial state of a Lindblad evolution obtained on a quantum device by implementing its time reversed master equation \cite{DoMu2022arx}. Concretely, we consider a master equation for a qubit weakly coupled to a thermal reservoir $\frac{d}{dt}\rho_t=\mathcal{L}_t(\rho_t)$ with
\begin{align}\label{eq:therm_for}
    \mathcal{L}_t(\rho_t) =&-i\omega[\sigma_z/2,\rho_t] +\gamma e^{\beta\omega}(\sigma_-\rho\sigma_+\frac{1}{2}\{\sigma_+\sigma_-,
    \rho_t\})\nonumber \\
    &+\gamma(\sigma_+\rho\sigma_-\frac{1}{2}\{\sigma_-\sigma_+,\rho_t\})
\end{align}
and its time reversed evolution  $\frac{d}{dt}\rho_t=-\mathcal{L}_t(\rho_t)$.

Thus, evolving a state for a time $t$ with \eqref{eq:therm_for} and then for a time $s\leq t$ with its time reversed evolution results in the state obtained by just evolving with \eqref{eq:therm_for} for a time $t-s$. We implement both the forwards and backwards evolution on the IBM device with our simulation scheme \eqref{eq:wittstock-paulsen2}.

In Fig. \ref{fig:recovery} data points for $t\leq t^*$ and various initial states reflect the thermalization dynamics of \eqref{eq:therm_for}. At $t=t^*$ the recovery sets in to approach earlier states in the dissipative time evolution. Note that the recovery is performed by mapping the state $\rho(t=t^*)$ directly to each recovered state with only one algorithm run per state. 

{\em Outlook.-} We have shown that the class of general qubit dynamical maps can be straightforwardly simulated using just four extremal CP maps each consisting of two pairs of Kraus operators. Environmental noise on a quantum system is generally described by a CP map. As the inverse of a CP map is a general dynamical map, one of the promising applications of our simulation scheme is to perform error mitigation by recovering the typically unknown bare qubit state in absence of any environmental coupling. We prove the viability of this in Fig. \ref{fig:recovery} on an IBM quantum processor. A next step is to determine the noise CP map of a single qubit of a quantum processor and to revert it using our simulation scheme and thus performing genuine quantum error mitigation.

\paragraph{Acknowledgements}
We thank P. Muratore-Ginanneschi, M. Donvil and J. Stockburger for valuable discussions. Financial support through the WM-BW within the Quantum Computing Competence Network BW (SiQuRe), the BMBF within QSens (QComp), and QSolid (BMBF) is gratefully acknowledged.

\bibliography{lit}{} 
\bibliographystyle{apsrev4-2}
\newpage

\onecolumngrid

\section{Supplemental Material}
The authors of \cite{WaBe2013} present a simple simulation scheme for completely positive trace preserving qubit maps, i.e. qubit channels. Recently, the protocol was realised experimentally by \cite{HaCa2021}. The method of \cite{WaBe2013} relies on an earlier mathematical results by \cite{RusSza2002,KingRus2001} which allow to write qubit channels as the convex sum of extremal channels. Concretely, these channels consist of two unitary transformations and the sum of two rather simple Kraus operators.

We are concerned with trace preserving general qubit dynamical maps. These maps are trace-preserving, self-adjoint but not necessarily positivity-preserving. Importantly, these maps can always be written as the difference of two completely positive maps. We show that here that any general dynamical map can be written as the weighted difference of two quantum channels. Combining this with the results of \cite{RusSza2002,KingRus2001,WaBe2013} we find a general simulation method for general qubit dynamics which needs just one ancilla qubit.

\section{General dynamical maps}
Completely bounded maps are linear maps $\Lambda$ for which the trivial extensions $\Lambda_N = \Lambda\otimes\mathbb{I}_N$ to larger Hilbert spaces  $\mathcal{H}\otimes\mathbb{C}_N$ satisfy
\begin{equation}
\|\Lambda\|_{cb}= \sup_N \|\Lambda_N\|<\infty.
\end{equation}
All maps linear maps acting on finite dimensional systems are completely bounded \cite{Paulsen2003}. When these linear maps are trace preserving and self-adjoint we call them general dynamical maps.
A physically relevant example of general dynamical maps are solutions to general time local master equations
\begin{equation} \label{eq:genTimeLoc}
\frac{d}{dt}\rho_S(t) = -i[H,\rho_S(t)] + \sum_k \gamma_k(t)\left(L_k\rho_S(t)L_k^\dagger - \frac{1}{2}\{L_k^\dagger L_k, \rho_S(t) \}\right)
\end{equation}
where the weights $\gamma_k(t)$ have no positivity requirements.

The Wittstock-Paulsen decomposition for completely bounded maps states that any completely bounded map can be written as the difference of two completely positive maps. Concretely, for every completely bounded $\Lambda$ there exist two completely positive maps $\Lambda_\pm$ such that
\begin{equation}\label{eq:decompotision}
\Lambda = \Lambda_+ - \Lambda_-.
\end{equation}
where $\|\Lambda_+\|,\, \|\Lambda_-\|\leq \|\Lambda\|_{cb}$.
We can write the completely positive maps in terms of Kraus operators 
\begin{equation}
\Lambda_+\rho = \sum_k M_k\rho M_k^\dagger,\quad \Lambda_-\rho = \sum_l K_l\rho K_l^\dagger.
\end{equation}
If $\Lambda$ is trace preserving then
\begin{equation}
\sum_k M_k^\dagger M_k - \sum_j K_j^\dagger K_j = \mathbb{I}.
\end{equation}
Since $\|\Lambda_-\|$ is bounded, there exists and $\alpha\geq\|\Lambda\|_{cb}$ such that 
\begin{equation}
\sum_j K_j^\dagger K_j\leq p \mathbb{I}
\end{equation}
Therefore the difference $p\mathbb{I}-\sum_j K_j^\dagger K_j =D^\dagger D$ is a positive matrix and 
\begin{equation}
\sum_j K_j^\dagger K_j + D^\dagger D=p \mathbb{I}.
\end{equation}
Let us now rewrite \eqref{eq:decompotision} as 
\begin{align}\label{eq:decomp}
\Lambda (\rho) = ({1+p})\left(\frac{\Lambda_+(\rho) + D\rho D^\dagger}{1+p}\right)  - p \left(\frac{\Lambda_-(\rho) + D\rho D^\dagger}{p}\right)
\end{align}
where the above equation is the weighted difference of completely positive maps. Moreso, the operators between the brackets are both trace 1.
\subsection{Decomposition into completely positive maps}
Let $\Sigma$ be a general dynamical map acting on a finite dimensional space. We compute the Choi matrix $C(\Sigma)$
\begin{align*}
C(\Sigma)= \sum_{ij} e_{ij} \otimes \Sigma(e_{ij}),
\end{align*}
where $e_{ij}$ are the elementary matrices.
The map $\Sigma$ can be obtained from the Choi matrix by
\begin{align*}
\Sigma(\rho)= \textrm{tr}_1(C(\Sigma)\, \rho^\top\otimes \mathbb{I})
\end{align*}
where $\textrm{tr}_1$ and $\rho^\top$ is the transpose of $\rho$.

 The Choi matrix has the property that its positivity is equivalent to the complete positivity of underlying map. Since $\Sigma$ is self-adjoint, $C(\Sigma)$ is a self-adjoint matrix and therefore diagonalisable. Let its eigenvectors and eigenvalues be $v_i$ and $\lambda_i$, we the define 
\begin{align*}
C_\pm(\Sigma) = \sum_i \max(0,\pm\lambda_i)\, v_iv_i^\dagger
\end{align*}
such that
\begin{equation}
C(\Sigma)=C_+(\Sigma)-C_-(\Sigma).
\end{equation}
We then define $\Lambda_\pm(\rho)$ in equation \eqref{eq:decompotision} as $\Lambda_\pm(\rho)=\textrm{tr}_1(C_\pm(\Sigma)\, \rho^\top\otimes \mathbb{I})$.

\section{Decomposition in extremal maps}
The authors of \cite{RusSza2002,KingRus2001} proved that any single qubit channel $\Lambda$ can be written as the convex sum of two channels "belonging to the closure of the set of extreme points of single qubit channels".

Let $\Lambda$ be a completely positive trace preserving map acting on qubit states. We can represent the action of $\Lambda$ on a state $\rho=\frac{1}{2}(\mathbb{I}+w\cdot \sigma)$ in terms of a $4\times 4$ matrix $\mathbb{T}$
\begin{equation}
\mathbb{T} = \begin{pmatrix}
1 & 0\\
t & T
\end{pmatrix},\quad \mathbb{T}_{ij} = \frac{1}{2}\text{tr}(\sigma_i\Lambda(\sigma_j))
\end{equation}
such that
\begin{equation}
\Lambda(\rho) = (\mathbb{I}+(t+ Tw) \cdot \sigma).
\end{equation}
There then exist two unitaries $U$, $V$ and a completely positive map $\Lambda'$ with diagonal $T$ such that 
\begin{equation}
\Lambda(\rho) = U\Lambda'(V\rho V^\dagger)U^\dagger
\end{equation}
where 
\begin{equation}
T' = \begin{pmatrix}
\lambda_1 & 0 & 0\\
0 & \lambda_2 &0 \\
0 & 0 & \lambda_3
\end{pmatrix}
\end{equation}
For the extremal maps, $\mathbb{T}'$ can be written as
\begin{equation}
\mathbb{T}' = \begin{pmatrix}
1 & 0 & 0 & 0 \\
0 & \cos \nu & 0 & 0\\
0 & 0 & \cos \mu & 0\\
\sin \mu\, \cos \mu & 0 &0 & \cos\mu \, \cos \mu
\end{pmatrix}
\end{equation}
The map defined by $\mathbb{T}'$ can be obtained from just 2 Kraus operators
\begin{equation}\label{eq:extreme-Kraus}
 F_0 = \begin{pmatrix}
 \cos \frac{\mu-\nu}{2}&0
 \\0 & \sin\frac{\mu+\nu}{2}
 \end{pmatrix},
 \qquad
  F_1 = \begin{pmatrix}
 0&\cos \frac{\mu+\nu}{2}
 \\ \sin\frac{\mu-\nu}{2}&0
 \end{pmatrix},
\end{equation}
By \cite{RusSza2002,KingRus2001} any completely positive qubit channel $\Lambda$ can be written as the convex sum of two convenient completely positive maps
\begin{equation}
\Lambda = \Lambda_1 + \Lambda_2.
\end{equation}
with
\begin{equation}\label{eq:ext-channel}
\Lambda_{j}(\rho) = U_j (F_{0,j} \,V_j\rho V_j^\dagger\,  F^\dagger_{0,j}+F_{1,j}\,V_j\rho V_j^\dagger\, F^\dagger_{1,j}) U_j^\dagger
\end{equation}
where $V_j$ and $U_j$ are unitaries and the Kraus operators $F_{0,j}$ and $F_{1,j}$.

\section{Completely Positive Maps}\label{app:CP}

\subsection{Diagonal representation of completely positive maps}\label{app:diag}
The action of a qubit channel on a qubit state can be expressed in terms of a 3 dimensional vector $t$ and a $3\times3$ matrix $T$. Let $\rho=\frac{1}{2}(\mathbb{I}+w\cdot \sigma)$, then
\begin{align*}
\Lambda(\rho) = \frac{1}{2}(\mathbb{I}+(t+ Tw) \cdot \sigma).
\end{align*}
In this section, we follow \cite{KingRus2001} and show how to find a diagonal representation. We write the singular value decomposition for $T$
\begin{equation}
T = V D W^\dagger
\end{equation}
where, since $T$ is a real matrix, $V$ and $W$ can be chosen to be real and thus orthogonal matrices. If $\det(V)=1$, $V$ is a rotation matrix, if $\det(V)=-1$ then $-V$ is a rotation matrix. Thus let $R_{1,2}$ are rotation matrices, then 
\begin{equation}
T = \frac{V}{\det V} \frac{D}{(\det V)(\det W^\dagger)} \frac{W^\dagger}{\det(W^\dagger)}
\end{equation}

\subsubsection{Rotation on the Bloch Sphere as a Unitary Transformation}
A rotation on the Bloch sphere around the axis $\hat{n}$ with and angle $\theta$ can be represented by the unitary transformation on the Hilbert space
\begin{align*}
U = \exp \left(-i\theta \,\hat{n}\cdot \frac{1}{2} \vec{\sigma} \right)
\end{align*}
We find the axis of a rotation matrix $R$ by solving 
\begin{equation}
(R-\mathbb{I})\hat{n} =0
\end{equation}
with $\|\hat{n}\|=1$ and the angle of rotation by
\begin{equation}
\text{tr}(R) = 1+ 2 \cos\theta
\end{equation}
\subsubsection{Diagonal representation}
We can thus write
\begin{align*}
\Lambda(\rho) = U_1 \Lambda_D(U_2 \rho U_2^\dagger) U_1^\dagger
\end{align*}
where 
\begin{align*}
\Lambda_D \left(\frac{1}{2}(\mathbb{I}+w \cdot \sigma)\right) = \frac{1}{2}\left(\mathbb{I}+\left(\frac{V^\dagger}{\det V^\dagger}\,t+ \frac{D}{(\det V)(\det W^\dagger)}w\right) \cdot \sigma\right)
\end{align*}
\subsection{Convex sum}
I repeat here the main results of \cite{RusSza2002} to show how a completely positive map can be written as the convex sum of two extremal maps.

Let $\Phi$ be a completely-positive trace-preserving qubit map and let $\hat{\Phi}$ be its adjoint. The Choi representation of $\hat{\Phi}$ is then a $4 \times 4$ matrix of the form 
\begin{align*}
C(\hat{\Phi}) =\begin{pmatrix}
\hat{\Phi}(E_{11}) &\hat{\Phi}(E_{12})  \\  \hat{\Phi}(E_{21})  &\hat{\Phi}(E_{22}) 
\end{pmatrix} =  \begin{pmatrix}
A & C \\ C^\dagger & \mathbb{I}-A
\end{pmatrix}.
\end{align*}
where $A$ and $C$ are $2\times 2$ matrices. The diagonal elements sum to one since $\hat{\Phi}(E_{11}) +\hat{\Phi}(E_{22}) = \hat{\Phi}(\mathbb{I}) = \mathbb{I} $ by trace preservation. Furthermore, since $\hat{\Phi}(E_{11}),\, \hat{\Phi}(E_{22})\geq 0$, we have that  $A\leq \mathbb{I}$.
Note that the Choi matrix of $\Phi$ and $\hat{\Phi}$ are related by 
\begin{align*}
C(\hat\Phi)= \overline{U^\dagger_{23} C({\Phi})U_{23}},
\end{align*}
where $U_{23}$ is the unitary matrix
\begin{align*}
U_{23}=U^\dagger_{23} = \begin{pmatrix}
1 & 0 & 0 & 0 \\
0 & 0 & 1 & 0 \\
0 & 1 & 0 & 0 \\
0 & 0 & 0 & 1
\end{pmatrix}.
\end{align*}
Such that $C(\hat\Phi)$ is positive definite if and only if $C(\Phi)$ positive definite is.

\begin{lem}
A matrix 
\begin{align*}
\begin{pmatrix}
A & C \\ C^\dagger & B
\end{pmatrix}
\end{align*}
is positive semi-definite if and only if $A\geq 0$, $B\geq 0$ and $C = \sqrt{A}R\sqrt{B}$ where $R$ is a contraction (i.e. $\|R\|\leq 1$).
\end{lem}
The following proposition then tells us something about the $R$ for generalised extreme points
\begin{prop}
A map is a generalised extreme point if and only if $C(\hat\Phi)$ is of the form
\begin{align*}
C(\hat\Phi) = 
\begin{pmatrix}
A & \sqrt{A}U\sqrt{\mathbb{I}-A} \\ \sqrt{\mathbb{I}-A}U^\dagger\sqrt{A}  & \mathbb{I}-A
\end{pmatrix}
\end{align*}
where $U$ is a unitary and $0\leq A\leq \mathbb{I}$.
\end{prop}

\begin{lem}
Let $R$ be a $2\times 2$ contraction, its singular value decomposition is of the form
\begin{align*}
R &= V \begin{pmatrix}
\cos \theta_1 & 0 \\ 0 & \cos \theta_2
\end{pmatrix} W^\dagger \\
&=\frac{1}{2} V \begin{pmatrix}
e^{i\theta_1} & 0 \\ 0 & e^{i\theta_2}
\end{pmatrix} W^\dagger +\frac{1}{2} V \begin{pmatrix}
e^{-i\theta_1} & 0 \\ 0 & e^{-i\theta_2}
\end{pmatrix} W^\dagger \\
& = \frac{U_1}{2} +\frac{U_2}{2} 
\end{align*}
where $U_{1,2}$ are unitaries.
\end{lem}
\begin{prop}
The Choi representation of the adjoint of any qubit channel $\Phi$ can be written as the convex sum of the Choi representation of two generalised extremal channels.
\begin{proof}
We have that
\begin{align*}
C(\hat\Phi) &= 
\begin{pmatrix}
A & \sqrt{A}R\sqrt{\mathbb{I}-A} \\ \sqrt{\mathbb{I}-A}R^\dagger\sqrt{A}  & \mathbb{I}-A
\end{pmatrix}\\
&=\frac{1}{2}\begin{pmatrix}A & \sqrt{A}U_1\sqrt{\mathbb{I}-A} \\ \sqrt{\mathbb{I}-A}U_1^\dagger\sqrt{A}  & \mathbb{I}-A
\end{pmatrix}+
\frac{1}{2}\begin{pmatrix}A & \sqrt{A}U_2\sqrt{\mathbb{I}-A} \\ \sqrt{\mathbb{I}-A}U_2^\dagger\sqrt{A}  & \mathbb{I}-A
\end{pmatrix}
\end{align*}
\end{proof}
\end{prop}

\section{Error Mitigation}

In order to obtain meaningful results from state-of-the-art NISQ quantum devices, it is necessary to employ methods aimed at mitigating the effect of noise. Among the various methods available in the field of quantum error correction and mitigation, we chose to use three methods to reduce the impact of device errors in our measurements. 

The first is to transpile, i.e. the operation that translates any theoretically designed circuit into the base of gates that the quantum device can actually implement, the circuit we have designed in order to reduce errors. This can be done, for example, by reducing the amount of CX gates needed to perform the given task.

The second is to choose from the set of qubits available in the quantum processor those that, at the time of each simulation, have the best coherence properties and protection from measurement errors with regard to the circuit we want to run. This can be done using specially designed functions from the Qiskit package, which can retrieve the state of each qubit in the processor directly from IBM's servers.

The third and slightly most refined method is readout error mitigation. Ideal measurements can be described by projection operators: each possible measurement result $i$ corresponds to a projection operator $P_i$, where $\sum_i P_i=I$. Performing a measurement on a quantum system in state $\rho$, the probability $p_i$ of obtaining result $i$ is given by $p_i={\rm Tr}(P_i\rho)$. Let us now assume that the measurement contains errors of the following kind: with probability $p_{1|0}$, the result `0` is turned into `1`, and vice versa $p_{0|1}$. Although the resulting measurement is no longer projective, the basic formalism (POVM: positive operator valued measure) remains the same - except for the fact that the operators become

\begin{equation*}
    P_0=\left(\begin{array}{cc}1-p_{1|0} & 0 \\ 0 & p_{0|1}\end{array}\right),\,  P_1=\left(\begin{array}{cc}p_{1|0} & 0 \\ 0 & 1-p_{0|1}\end{array}\right)
\end{equation*}

The above operators are not projection operators but instead positive, self-adjoint. An error of the above kind (which only concerns the assignment of measurement results) can be corrected by classical post-processing of measurement results using a method called \textit{LocalReadoutError} implemented in the Qiskit experiments library.
\end{document}